\documentclass[nofootinbib,twocolumn,prd,groupedaddress,preprintnumbers]{revtex4-1}
\usepackage[english]{babel}
\usepackage{slashed}
\usepackage{bm}
\usepackage{graphicx}
\usepackage{color}
\usepackage{upgreek}
\usepackage{amssymb,amsmath}
\usepackage{mathrsfs}
\usepackage[normalem]{ulem}
\usepackage{bbm}
\usepackage{array}
\usepackage[usenames,dvipsnames]{xcolor}
\usepackage[utf8]{inputenc}
\RequirePackage[colorlinks=true,urlcolor=Magenta,anchorcolor=blue,citecolor=blue,filecolor=blue,
linkcolor=Magenta,menucolor=blue,linktocpage=true,pdfproducer=medialab]{hyperref}
\usepackage{catchfile} 
\newcommand{\getenv}[2][]{%
  \CatchFileEdef{\temp}{"|kpsewhich --var-value #2"}{}%
  \if\relax\detokenize{#1}\relax\temp\else\let#1\temp\fi}
\getenv[\USER]{USER}
\def\nn{\nonumber}

\def\hc{\text{h.c.}}

\def\derp{\partial}

\newcommand{\vev}[1]{\langle#1\rangle}
\newcommand{\ov}[1]{\overline{#1}}
\def\be{\begin{equation}}
\def\ee{\end{equation}}
\newcommand{\TeV}{\;\text{TeV}}
\newcommand{\GeV}{\;\text{GeV}}
\newcommand{\MeV}{\;\text{MeV}}
\newcommand{\keV}{\;\text{keV}}
\newcommand{\eV}{\;\text{eV}}
\newcommand{\m}{\;\text{m}}
\newcommand{\mm}{\;\text{mm}}
\newcommand{\fb}{\;\text{fb}}
\def\cG{\mathcal{G}}
\def\cH{\mathcal{H}}
\def\cM{\mathcal{M}}

\def\sL{\mathscr{L}}
\def\tf{\mathtt{f}}
\def\s{\sigma}
\def\lsp{\lambda_{s\phi}}
\def\UPQ{U(1)_\text{PQ}}
\newcommand{\blue}[1]{\color{blue} #1 \color{black} }

\begin{document}

\title{\boldmath{\blue{Testable Axion-Like Particles In The Minimal Linear $\sigma$ Model}}}

\author{{\bf J. Alonso-Gonz\'alez}}
\email{j.alonso.gonzalez@csic.es}
\affiliation{\vspace{1mm} 
Departamento de F\'isica Te\'orica and Instituto de F\'isica Te\'orica, IFT-UAM/CSIC, Universidad Aut\'onoma de Madrid, 
Cantoblanco, 28049, Madrid, Spain}
\author{{\bf L. Merlo}}
\email{luca.merlo@uam.es}
\affiliation{\vspace{1mm} 
Departamento de F\'isica Te\'orica and Instituto de F\'isica Te\'orica, IFT-UAM/CSIC, Universidad Aut\'onoma de Madrid, 
Cantoblanco, 28049, Madrid, Spain}
\author{{\bf F. Pobbe}}
\email{federico.pobbe@pd.infn.it}
\affiliation{\vspace{1mm}
Dipartimento di Fisica e Astronomia, Universit\`a di Padova and INFN, Sezione di Padova, via Marzolo 8, I-35131 Padova, Italy}
\author{{\bf S. Rigolin}}
\email{stefano.rigolin@pd.infn.it}
\affiliation{\vspace{1mm}
Dipartimento di Fisica e Astronomia, Universit\`a di Padova and INFN, Sezione di Padova, via Marzolo 8, I-35131 Padova, Italy}
\author{{\bf O. Sumensari}}
\email{olcyr.sumensari@pd.infn.it}
\affiliation{\vspace{1mm}
Dipartimento di Fisica e Astronomia, Universit\`a di Padova and INFN, Sezione di Padova, via Marzolo 8, I-35131 Padova, Italy}

\begin{abstract}
Axion and axion-like particle models are typically affected by a strong fine-tuning problem in conceiving the electroweak and the Peccei-Quinn breaking scales. Within the context of the Minimal Linear $\sigma$ Model, axion-like particle constructions are identified where this hierarchy problem is solved, accounting for a TeV Peccei-Quinn breaking scale and a pseudoscalar particle with a mass larger than $10$ MeV. Potential signatures at colliders and B-factories are discussed. 
\end{abstract}
 \preprint{FTUAM-18-18}
 \preprint{IFT-UAM/CSIC-18-68}

\maketitle

%
%%%%%%%%%%%%%%%%%%%%%%%       I        %%%%%%%%%%%%%%%%%%%%%
%
\section{Introduction}
\label{Sect:Intro}

A suggestive mechanism to protect the Higgs mass from radiative corrections arises when the Higgs field belongs to the Nambu 
Goldstone Boson (GB) sector of a model equipped with a global symmetry $\cG$ spontaneously broken, by an unknown strongly 
interacting dynamics, to a subgroup $\cH$. In the Composite Higgs (CH) framework, the Standard Model (SM) GBs and the Higgs field 
parametrise (some of) the coset $\cG/\cH$ coordinates and are forced to be strictly massless \cite{Kaplan:1983fs,Kaplan:1983sm,
Banks:1984gj,Dugan:1984hq}. The gauging of the SM symmetries and the introduction of fermionic Yukawa couplings introduce an explicit 
breaking of $\cG$, leading to a non-vanishing mass term for the Higgs and to the spontaneous breaking of the electroweak (EW) symmetry. 

The minimal CH model (MCHM)~\cite{Agashe:2004rs}, is based on the symmetric coset $\cG/\cH=SO(5)/SO(4)$. Extended constructions have been 
presented in Refs.~\cite{Gripaios:2009pe, Mrazek:2011iu}. CH models are typically written in the language of effective field theories, 
parametrising the lack of knowledge of the strongly interacting sector with a large set of unknown coefficients. This description is 
consequently valid only up to a scale $\Lambda_s$, where strong dynamics resonances are supposed to appear. 

The Minimal Linear $\sigma$-Model (ML$\s$M)~\cite{Barbieri:2007bh,Gertov:2015xma,Feruglio:2016zvt,Gavela:2016vte} is, instead, a renormalisable 
model that represents a convenient and well-defined framework that, at need, by integrating out the extra scalar degree of freedom (dof) $\sigma$, 
matches the usual effective non-linear MCHM~\cite{Agashe:2004rs,Alonso:2014wta,Panico:2015jxa,Hierro:2015nna} Lagrangian, or the more general 
Higgs Effective Field Theory Lagrangian~\cite{Feruglio:1992wf,Grinstein:2007iv,Contino:2010mh,Alonso:2012px,Alonso:2012jc,Alonso:2012pz,
Buchalla:2013rka,Brivio:2013pma,Brivio:2014pfa,Gavela:2014vra,Gavela:2014uta,Alonso:2015fsp,Brivio:2015kia,Gavela:2016bzc,Alonso:2016btr,Eboli:2016kko,
Brivio:2016fzo,Alonso:2016oah,deFlorian:2016spz,Merlo:2016prs,Hernandez-Leon:2017kea,Alonso:2017tdy,Kozow:2019txg}. Following the treatment of 
Ref.~\cite{Feruglio:2016zvt}, the symmetry content of the ML$\s$M consists of a global $SO(5)$ spontaneously broken to $SO(4)$ when the 
scalar $SO(5)$ quintuplet $\phi$, containing the SM EW doublet Higgs $H$ and the EW singlet $\sigma$, acquires a non-vanishing vacuum 
expectation value (VEV) $f$, assumed to be at the TeV scale in order to solve the Higgs naturalness problem. Besides the SM gauge bosons and fermions, the spectrum accounts for additional sets of vector-like exotic fermions, both in the trivial and in the fundamental 
representations of $SO(5)$, in such a way that $SO(5)$ invariant Yukawa couplings can be introduced. Moreover, non-vanishing masses for 
the SM fermions originate through the partial-compositeness mechanism~\cite{Kaplan:1991dc,Contino:2004vy}, by adding bilinear ($SO(5)$ explicit breaking) operators between the SM and exotic fermions. Finally, the symmetry sector is customarily enlarged by an extra $U(1)_X$ symmetry 
to correctly account for the SM hypercharge assignment. By extending the ML$\s$M spectrum with an additional complex scalar field $s$, singlet 
under SM and $SO(5)$ symmetries, and by supplementing it with an additional global Abelian symmetry {\it \`a la} Peccei-Quinn 
(PQ)~\cite{Peccei:1977hh}, $\UPQ$, an axion or an axion-like particle (ALP) can also be introduced. Such a framework has been dubbed 
Axion-ML$\s$M (AML$\s$M)~\cite{Brivio:2017sdm,Merlo:2017sun}. 

The tree-level renormalisable scalar potential associated to the AML$\s$M, describing the spontaneous $SO(5)/SO(4)$ and PQ symmetry 
breaking, reads~\cite{Merlo:2017sun} 
\be
\begin{split}
V(\phi,\,s)=&\lambda(\phi^T\phi-f^2)^2+\lambda_s(2s^*s-f_s^2)^2+\\
&-2\lsp(s^*s)(\phi^T\phi)+\ldots\,,
\end{split}
\label{ScalarPotential}
\ee
where $\lambda$, $\lambda_s$ and $\lsp$ are dimensionless parameters, $f$ and $f_s$ the $SO(5)$ and $\UPQ$ symmetry breaking scales, 
and the dots stand for all possible $SO(5)$ and/or PQ explicit breaking terms necessary to guarantee the EW symmetry breaking, a 
viable SM spectrum and the renormalisability of the model. 

It is customary to parametrise the complex scalar singlet $s$, in the PQ symmetry broken phase, with an exponential notation,
\be
s=\frac{v_r+r}{\sqrt{2}} e^{i\,a/f_a}\,,
\ee
with $r$ the radial component field and $a$ the pseudoscalar field, to be identified with the axion or ALP dof. The axion (or ALP) 
decay constant $f_a \equiv\vev{r}\equiv v_r$ is typically of the order of the PQ breaking scale $f_s$ and may undergo strong 
constraints arising from the experimental limits on the pseudoscalar coupling to photons. In the case of $a$ being the QCD axion, 
with mass $m_a < 10\eV$, the bound\footnote{A recent analysis~\cite{Giannotti:2017hny} shows a preferred region of the parameter space 
$g_{a\gamma\gamma}\times g_{aee}$, being the latter the effective coupling of the axion with two electrons. When interpreted in terms of 
the DFSZ~\cite{Zhitnitsky:1980tq,Dine:1981rt} or KSVZ~\cite{Kim:1979if,Shifman:1979if} axion models, the best fit point is in the border 
of the perturbative unitarity of the Yukawas, while the fit is inconclusive at more than $2\sigma$. This region will be tested by future 
ARIADNE~\cite{Arvanitaki:2014dfa} and IAXO~\cite{Armengaud:2014gea} experiments.} on the scale $f_a$ is~\cite{Jaeckel:2015jla,
Anastassopoulos:2017ftl,Bauer:2017ris} 
\be
f_a\gtrsim|g_{a\gamma\gamma}|\times10^7\GeV\,, 
\label{faBound}
\ee
where $g_{a\gamma\gamma}$ is the adimensional effective coupling of the axion to two photons and depends on the fermionic spectrum 
and on the PQ charge assignment considered. For masses $10\eV < m_a < 0.1\GeV$ the constraints become even stronger~\cite{Millea:2015qra}.

The bound in Eq.~(\ref{faBound}) strongly crashes with the requirement of a natural EW scale. Indeed, as explicitly shown in 
Ref.~\cite{Merlo:2017sun}, either the coupling $\lsp$ in Eq.~(\ref{ScalarPotential}) is unnaturally set to 0, or the ``effective'' 
$SO(5)/SO(4)$ breaking scale, labelled $f_R$, runs to the highest scale, $f_R\approx f_s \approx f_a$, reintroducing a strong fine-tuning 
between the EW and the CH scale, $\xi = v^2/f^2_R \ll 1$. This suggests that the AML$\s$M framework is ``natural'' only if $f_s$, and 
therefore $f_a$, are at the TeV scale.

To escape the constraint in Eq.~(\ref{faBound}), two approaches can be outlined. The first is still to rely on the QCD axion paradigm, as the solution to the strong CP problem: in this case a specific ({\it ad hoc}) choice of the PQ charges can be identified such that the $a\gamma\gamma$ coupling is vanishing~\cite{Craig:2018kne} and consequently the astrophysical constraints are automatically evaded. The second approach consists in abandoning, partially or completely, the QCD ansatz and considering, instead, an ALP particle: then the inverse proportionality relation between the QCD axion mass and its decay constant does not need to be enforced anymore and a mass larger than $0.1\GeV$ can be achieved, relaxing the astrophysics bounds on the $a\gamma\gamma$ coupling, $f_a\gtrsim |g_{a\gamma\gamma}|\GeV$. 

In Ref.~\cite{Merlo:2017sun}, a minimal ALP scenario in the AML$\s$M framework has been considered, assuming that the PQ dynamics does not intervene in the explicit breaking of the $SO(5)$ symmetry, i.e. the two scales $f$ and $f_s$ are independent. The scale $f$ can be taken in the TeV range and the phenomenology associated to the $SO(5)/SO(4)$ sector turns out to be very similar to the one described in the original ML$\s$M in Ref.~\cite{Feruglio:2016zvt}; the scale $f_a$, and therefore $f_s$, can also be taken in the TeV range, opening the possibility to test this model both at colliders and at B-factories. Moreover, no fine-tuning between the two scales $f$ and $f_s$ is necessary in this model.

In this letter, the mechanisms behind the PQ and the $SO(5)$ symmetry breaking are instead identified assuming $f_s = f$ around the 
TeV scale (the possibility of $f_s \approx f$ has already been considered in Ref.~\cite{Brivio:2017sdm}, where however only the QCD axion scenario has been investigated, with $f_s \approx f\gg\TeV$). This is obtained by substituting the scales appearing in the $SO(5)$ explicit breaking terms with the singlet scalar $s$: once this field develops a VEV, these terms break $SO(5)$ spontaneously and not explicitly, thus linking $f$ with $f_s$. In this context, where $SO(5)$ is dynamically broken, alternative constructions with respect to the ALP solution considered in Ref.~\cite{Merlo:2017sun} can be considered. Moreover, as shown in Sect.~\ref{Sect:ALP}, these AML$\s$M realisations can be testable at colliders and flavour factories.

In Sec.~\ref{Sect:MLsM} the introduction of an ALP in the ML$\s$M is reviewed, while in Sec.~\ref{Sect:MLsMminimal} viable, natural and 
minimal models are identified. The associated phenomenology is discussed in Sec.~\ref{Sect:ALP} and final remarks are deferred to 
Sec.~\ref{Sect:Conc}. In the App.~\ref{APP:PQTrans}, the ALP Lagrangian is reported with the explicit expression for the ALP-gauge boson couplings. 

%
%%%%%%%%%%%%%%%%%%%%%%%       II        %%%%%%%%%%%%%%%%%%%%%
%

\boldmath
\section{Introducing $\UPQ$ in the ML$\sigma$M}
\label{Sect:MLsM}
\unboldmath

The AML$\s$M~\cite{Brivio:2017sdm,Merlo:2017sun} is characterised by the global symmetry $SO(5)\times U(1)_X\times \UPQ$. The spectrum includes 
the SM gauge bosons, the $SO(5)$ scalar quintuplet $\phi$, the complex scalar $s$, the third family SM fermions and four exotic fermions, two $SO(5)$ quintuplets $\psi^{_{(')}}$ and two $SO(5)$ singlets $\chi^{_{(')}}$. Only the exotic fields transform under $U(1)_X$, with the unprimed (primed) fields having a charge $+2/3$ $(-1/3)$ respectively. The SM hypercharge $Y$ is given by the following combination between the generators of $SU(2)_R \subset SO(5)$ and $U(1)_X$: 
\be
Y=\Sigma_R^{(3)}+X\,.
\ee
The decomposition of the exotic fermions and their transformation properties under the SM gauge symmetry can be found in Tab.~\ref{DecompositionFermions}.

\begin{table}[h!]
\hspace{-5mm}
\begin{tabular}{l|c|c|c|c|}
			& $SO(5)\times U(1)_X$ & $SU(3)_C\times SU(2)_L\times U(1)_Y$ & $Q_\text{em}$ \\
\hline
& & & \\[-2mm]
		&				& $K=({\bf3},{\bf2},7/6)$	& $5/3, 2/3$ 	\\[1mm]
$\psi$ 	& $({\bf5},\, 2/3)$ 	& $Q=({\bf3},{\bf2},1/6)$ 	& $2/3, -1/3$ 	\\[1mm]
		&				& $T_5=({\bf3},1,2/3)$	& $2/3$ 		\\[1mm]
		\hline
& & & \\[-2mm]
$\chi$ 	& $(1,\, 2/3)$ 		& $T_1=({\bf3},1,2/3)$ 	& $2/3$ 		\\[1mm]
\hline
& & & \\[-2mm]
		&				& $Q'=({\bf3},{\bf2},1/6)$	& $2/3, -1/3$ 	\\[1mm]
$\psi'$ 	& $({\bf5},\, -1/3)$ 	& $K'=({\bf3},{\bf2},-5/6)$ 	& $-1/3, -4/3$ 	\\[1mm]
		&				& $B_5=({\bf3},1,-1/3)$	& $-1/3$ 		\\[1mm]
\hline
& & &\\ [-2mm]
$\chi'$ 	& $(1,\,-1/3)$ 	& $B_1=({\bf3},1,-1/3)$ 	& $-1/3$ 		\\[1mm]
\hline
\end{tabular}
\caption{\em Decomposition of the exotic fermions and their transformation properties under the SM gauge symmetry. \label{DecompositionFermions}}
\end{table}

Generalising the notation of Ref.~\cite{Brivio:2017sdm,Merlo:2017sun}, the \mbox{$SO(5)\times U(1)_X$} 
invariant Lagrangian containing the fermionic interactions can be written as follows:
\begin{widetext}
\begin{align}
\sL_{\rm f} =&\,\, 
\overline{q}_L i\slashed{D}\,q_L\,+\,\overline{t}_R i\slashed{D}\,t_R\,+\,\overline{b}_R i\slashed{D}\,b_R \,+ \overline{\psi} \left[i \slashed{D} - M_5 \right] \psi \,+\,\overline{\chi} \left[i \slashed{D} - M_1 \right]\chi \,+\overline{\psi'} \left[i \slashed{D} - M'_5 \right]\psi' \,+\,\overline{\chi'} \left[i \slashed{D} - M'_1 \right]\chi'\nn\\
&-\,\Big[ y_1\, \overline{\psi}_L\, \phi\, \chi_R + y_2\, \overline{\psi}_R\, \phi\, \chi_L +z_1\,\overline{\chi}_R\,\chi_L\,s + \tilde{z}_1\,\overline{\chi}_R\,\chi_L\,s^\ast + z_5\,\overline{\psi}_R\,\psi_L\,s + \tilde{z}_5\,\overline{\psi}_R\,\psi_L\,s^\ast +  \label{FermionLagrangian}\\
&\phantom{-\,\{}+\left(\Lambda_1+k_1\,s+\tilde{k}_1\,s^*\right) \left(\overline{q}_L \Delta_{2\times5} \right) \psi_R  
+  \left(\Lambda_2+k_2\,s+\tilde{k}_2\,s^*\right) \, \overline{\psi}_L \left (\Delta_{5\times1} t_R \right) 
+  \left(\Lambda_3+k_3\,s+\tilde{k}_3\,s^*\right) \,\overline{\chi}_L t_R + \hc \Big]\,+ \nn\\   
&-\,\Big[y'_1\, \overline{\psi'}_L\, \phi\, \chi'_R + y'_2\,\overline{\psi'}_R\, \phi\,\chi'_L+z'_1\,\overline{\chi'}_R\,\chi'_L\,s + \tilde{z}'_1\,\overline{\chi'}_R\,\chi'_L\,s^\ast + 
z'_5\,\overline{\psi'}_R\,\psi'_L\,s + \tilde{z}'_5\,\overline{\psi'}_R\,\psi'_L\,s^\ast+ \nn\\ 
&\phantom{-\,\{}+\left(\Lambda'_1 +k'_1\,s+\tilde{k}'_1\,s^*\right)\left(\overline{q}_L \Delta'_{2\times5} \right) \psi'_R  
+  \left(\Lambda'_2 +k'_2\,s+\tilde{k}'_2\,s^*\right) \,\overline{\psi'}_L \left (\Delta'_{5\times1} b_R \right) 
+  \left(\Lambda'_3 +k'_3\,s+\tilde{k}'_3\,s^*\right) \,\overline{\chi'}_L b_R + \hc \Big]\,. \nn
\end{align}
\end{widetext}
The first line contains the kinetic terms for all the fermions plus the direct mass terms, $M^{(\prime)}_{1,5}$, of the exotic fields. 
The second and third lines refer to the top sector. The terms proportional to $y_i$ are $SO(5)$ invariant Yukawa terms between exotic 
fermions and the scalar quintuplet. The ones proportional to $z_i$ and $\tilde{z}_i$ are Yukawa-type interactions, between the exotic 
fermions and the $SO(5)$ singlet scalar, which contribute to the exotic fermion masses once the PQ symmetry breaking occurs. In the 
third line the partial compositeness operators linking the exotic fermions and the top quark are included, needed for providing a 
non-vanishing top mass. The first two terms, which are proportional to $\Delta_{i\times j}$, explicitly break the $SO(5)$ symmetry, 
while the last one is, instead, $SO(5)$ preserving. The $\Delta_{i\times j}$ quantities play the role of spurions~\cite{DAmbrosio:2002vsn,Cirigliano:2005ck,Davidson:2006bd,Alonso:2011jd,Dinh:2017smk} 
for the $SO(5)\times U(1)_X$ symmetry. Finally the last two lines refer 
to the bottom sector and all the previous comments on the unprimed terms apply to their primed counterparts as well. 

With the exotic fermions acquiring masses larger than the EW scale, a fermionic Seesaw mechanism provides the masses for the SM fermions~\cite{Kaplan:1991dc,Contino:2004vy,Dugan:1984hq,Galloway:2010bp,Grinstein:2010ve,Feldmann:2010yp,Guadagnoli:2011id,Buras:2011zb,Buras:2011wi,Feldmann:2016hvo,Alonso:2016onw}: the Leading Order (LO) contribution reads:
\be
\begin{split}
m_t=&\dfrac{y_1\,\Lambda_1(v_r)\,\Lambda_3(v_r)\,v_h}{M_1(v_r)\,M_5(v_r)-y_1\,y_2\,(v_h^2+v_\s^2)}+\\
&-\dfrac{y_1\, y_2\, \Lambda_1(v_r)\,\Lambda_2(v_r)\,v_h\, v_\s}{M_1(v_r)\,M_5^2(v_r)-y_1\,y_2\, M_5(v_r)\,(v_r^2+v_\s^2)}\,,
\end{split}
\label{TopMass}
\ee
where $v_h$, $v_\s$ and $v_r$ are the VEVs of the physical field $h$, $\s$ and $r$, satisfying to $v_h^2+v^2_\sigma=f^2$, and the functions 
of $v_r$ are defined as
\be
\begin{aligned}
\Lambda_i(v_r)&\equiv \Lambda_i+(k_i+\tilde{k}_i)v_r\,,\\
M_i(v_r)&\equiv M_i+(z_i+\tilde{z}_i)v_r\,.
\end{aligned}
\ee
Similar expressions hold for the bottom sector too.

A general comment has to be highlighted. In the Lagrangian in Eq.~(\ref{FermionLagrangian}) and in the expressions for the SM fermion 
masses above, not all the terms can actually be present at the same time. Depending on the specific choice of the PQ charges, several 
terms are simply forbidden: in particular only one term among those proportional to $\Lambda_i$, to $k_i$ and to $\tilde{k}_i$ is 
allowed for a given PQ charge assignment; a similar observation holds for the terms proportional to $M_i$, $z_i$ and $\tilde{z}_i$. 

Once the fermionic Lagrangian is fully determined, the computation of the 1-loop contributions to the scalar potential is 
straightforward: the Coleman-Weinberg~(CW) formula~\cite{Coleman:1973jx} allows to extract the divergences generated at 1-loop 
with internal fermion and gauge boson lines. This aspect has been described in details in Ref.~\cite{Feruglio:2016zvt,Merlo:2017sun} 
for the ML$\s$M without and with the presence of the PQ symmetry. In general, several divergent contributions arise at one loop that 
cannot be re-absorbed in the tree-level $SO(5)$ invariant scalar potential in Eq.~(\ref{ScalarPotential}). In consequence, to have a 
renormalisable Lagrangian, consistent with a viable EW symmetry breaking, the corresponding terms need to be added to 
the tree-level scalar potential. As two is the minimum number of explicit $SO(5)$ breaking terms needed to have a viable EW breaking 
sector, constructions with only two extra parameters in Eq.~(\ref{ScalarPotential}) have been dubbed ``minimal''.

\boldmath
\section{Viable, Natural \& Minimal AML$\sigma$M}
\label{Sect:MLsMminimal}
\unboldmath

A proper model should be viable, natural and minimal. In order to construct an AML$\s$M satisfying these three features, the 
following guiding conditions are required: i) third generation SM fermion masses are generated at LO and therefore the 
expression in Eq.~(\ref{TopMass}) must not vanish; ii) no large hierarchy is present between the $SO(5)$ and $\UPQ$ breaking scales; 
iii) the model depends on the minimal possible number of parameters. 

\begin{table*}
\begin{tabular}{l|c||cccccccccc|ccc|}
& Conditions on PQ charges & $y_1$ & $y_2$ & $\Lambda_1$  & $k_1$ & $\tilde{k}_1$ & $\tilde{k}_3$ & $\tilde{z}_1$ & $M_5$ 
& $z_5$ & $\tilde{z}_5$ & $\Delta_\psi$ & $\Delta_\chi$ & $\Delta_t$ \\
\hline
& & & & & & & & & & & & & & \\[-2mm]
$\cM_1$	& $n_{\psi_L}=n_{\chi_R}=n_{\psi_R}=n_{\chi_L}- n_s=n_{q_L}=n_{t_R}-2n_s$
& \checkmark & & \checkmark & && \checkmark  & \checkmark  &  \checkmark & & & 0 & $n_s$ & $-2n_s$ \\[2mm]	
$\cM_2$	& $n_{\psi_L}=n_{\chi_R}=n_{\psi_R}=n_{\chi_L}- n_s=n_{q_L}+ n_s=n_{t_R}-2n_s$
& \checkmark & & &  & \checkmark & \checkmark  & \checkmark &  \checkmark & & & 0 & $n_s$ & $-3n_s$ \\[2mm]	
$\cM_3$	& $n_{\psi_L}=n_{\chi_R}=n_{\psi_R}+ n_s=n_{\chi_L}- n_s=n_{q_L}+2n_s=n_{t_R}-2n_s$
& \checkmark & & & & \checkmark & \checkmark & \checkmark & & & \checkmark & $n_s$ & $n_s$ & $-4n_s$ \\[2mm]
$\cM_4$	& $n_{\psi_L}=n_{\chi_R}=n_{\psi_R}+ n_s=n_{\chi_L}- n_s=n_{q_L}+ n_s=n_{t_R}-2n_s$
& \checkmark & & \checkmark & & & \checkmark  & \checkmark & & & \checkmark & $n_s$ & $n_s$ & $-3n_s$ \\[2mm]
$\cM_5$	& $n_{\psi_L}=n_{\chi_R}=n_{\psi_R}+ n_s=n_{\chi_L}- n_s=n_{q_L}=n_{t_R}-2n_s$
& \checkmark & & & \checkmark & & \checkmark  & \checkmark &  & & \checkmark & $n_s$ & $n_s$ & $-2n_s$ \\[2mm]
$\cM_6$	& $n_{\psi_L}=n_{\chi_R}=n_{\psi_R}- n_s=n_{\chi_L}- n_s=n_{q_L}- n_s=n_{t_R}-2n_s$
& \checkmark & \checkmark & \checkmark  && & \checkmark & \checkmark &  & \checkmark & & $-n_s$ & $n_s$ & $-n_s$ \\[2mm]
$\cM_7$	& $n_{\psi_L}=n_{\chi_R}=n_{\psi_R}=n_{\chi_L}- n_s=n_{q_L}- n_s=n_{t_R}-2n_s$
& \checkmark & & & \checkmark & & \checkmark  & \checkmark & \checkmark & & & 0 & $n_s$ & $-n_s$\\[2mm]
\hline	
\end{tabular}
\caption{\em List of the viable, natural and minimal AML$\s$M realisations, defined by the conditions on the PQ charges of the fermion 
fields written in terms of the charge $n_s$ of the PQ scalar field. The constants allowed in the Lagrangian are indicated with 
``\checkmark'', while all the remaining Lagrangian parameters that are not listed in this table are not allowed for  symmetry reasons. 
On the right side, the corresponding values for $\Delta_\psi$, $\Delta_\chi$ and $\Delta_t$ are 
listed.}
\label{TabModels}
\end{table*}

In order to identify the PQ charge assignments compatible with these three requirements, it is useful to 
introduce the following five PQ charges differences: for the top sector,
\be
\begin{gathered}
\Delta_{y_1}\equiv n_{\psi_L}-n_{\chi_R}\\
\begin{aligned}
&\Delta_{\Lambda_1}\equiv n_{q_L}-n_{\psi_R}\qquad
&&\Delta_{\Lambda_3}\equiv n_{\chi_L}-n_{t_R}\\
&\Delta_{\chi}\equiv n_{\chi_L}-n_{\chi_R}\qquad
&&\Delta_{\psi}\equiv n_{\psi_L}-n_{\psi_R}\,.
\end{aligned}
\end{gathered}
\label{DefDeltas}
\ee
Similar quantities can be defined for the bottom sector by replacing the unprimed fields with the primed ones.

Condition i) is satisfied by requiring that none among $y_1$, $\Lambda_1(v_r)$, $\Lambda_3(v_r)$, $M_1(v_r)$ and $M_5(v_r)$ is vanishing. 
Alternative possibilities with non-vanishing $\Lambda_2(v_r)$ turn out to be non-minimal. In terms of the quantities defined above, 
this corresponds to
\be
\begin{gathered}
\Delta_{y_1}=0\,,\qquad
\Delta_{\Lambda_1}=\{0,\,\pm n_s\}\,,\qquad
\Delta_{\Lambda_3}=\{0,\,\pm n_s\}\,,\\
\Delta_{\chi}=\{0,\,\pm n_s\}\,,\qquad
\Delta_{\psi}=\{0,\,\pm n_s\}\,. \nn
\end{gathered}
\ee
Whenever one of these quantities vanishes, the corresponding allowed term in the Lagrangian is the constant one: i.e. $y_1$, $\Lambda_{i}$ 
and $M_i$. On the other hand, if any of the charge differences is equal to $-n_s$ ($+n_s$), the corresponding term, allowed in the Lagrangian, 
is proportional to $s$ ($s^*$). As an example, $\Delta_{\chi}=0$ indicates that $\chi_L$ 
and $\chi_R$ transform under $\UPQ$ with the same charge and therefore the term $M_1\overline{\chi}_R\chi_L$ is invariant under $\UPQ$ and should 
be kept in the Lagrangian. If, instead, $\Delta_{\chi}=-n_s$, then the $z_1\,\overline{\chi}_R\,\chi_L\,s$ term is the invariant one. 
There are $3^4=81$ possible different configurations compatible with condition i) for a single fermion sector, while any value 
different from 0 or $\pm n_s$ leads to vanishing SM fermion masses.

The naturalness requirement, condition ii), is satisfied only if all the scales in the Lagrangian, except for $M_1$ and $\Lambda_3$, 
are in the TeV range. The $SO(5)$ and $\UPQ$ breaking scales $f$ and $f_s$ need to satisfy to this condition in order to avoid large 
fine-tunings in the tree-level scalar potential, as discussed in the Introduction. For the other quantities, such as $M_5$, $\Lambda_1$ 
and $\Lambda_2$, the reason resides in the fact that they correct the scalar potential parameters at one-loop (see Ref.~\cite{Merlo:2017sun} 
for details). If these parameters are much larger than the TeV, large fine-tunings would be necessary in order to guarantee a viable 
EW VEV. $M_1$ and $\Lambda_3$ evade this condition because they do not enter the CW contributions: as already pointed out in Ref.~\cite{Brivio:2017sdm}, 
they only need to satisfy $\Lambda_3/M_1\sim1$ in order to provide a viable value for the mass of third generation SM quarks -- see 
Eq.~(\ref{TopMass}) -- assuming natural Yukawa couplings $y_i$.

The minimality condition iii) only concerns the number of parameters that enter the scalar potential once considering the 1-loop 
contributions. Two divergent terms, proportional to $h^2$ and $h^4$, arise from the CW potential induced by the gauge bosons: 
these divergences are independent of the specific PQ charge assignment and therefore the corresponding terms necessarily enter the 
final scalar potential. Minimal constructions are those where the fermionic CW potential does not introduce any additional divergence 
that cannot be absorbed by a redefinition of the parameters in Eq.~(\ref{ScalarPotential}) or of $h^2$ or $h^4$, as discussed in 
Ref.~\cite{Merlo:2017sun}.

Out of the 81 possible AML$\s$M constructions, only seven satisfy to all three conditions and they are listed in Tab.~\ref{TabModels}, 
defined by the PQ charges of the fermion fields, written as a function of the charge $n_s$ of the PQ scalar field. In the same table, 
the parameters entering the Lagrangian are explicitly reported. On the right side, the corresponding values for $\Delta_\psi$, $\Delta_\chi$ and 
\be
\Delta_t\equiv n_{q_L}-n_{t_R}
\label{DefDeltat}
\ee
are listed, as they will be relevant in the phenomenological section that follows. A sibling for each configuration can be found by 
replacing $n_s\to-n_s$, $\tilde{k}_3\to k_3$, $z_1\to\tilde{z}_1$, $k_1\leftrightarrow\tilde{k}_1$ and $z_5\leftrightarrow\tilde{z}_5$. 
A charge assignment and its own sibling, for a given fermion sector, are completely equivalent. All the remaining Lagrangian parameters 
that are not listed in this table are not allowed for symmetry reasons. Similar considerations hold for the bottom quark sector, in terms 
of the PQ charge differences $\Delta_{\psi'}$, $\Delta_{\chi'}$ and 
\be
\Delta_b\equiv n_{q_L}-n_{b_R}\,.
\label{DefDeltab}
\ee

The top and bottom sectors are not completely independent as $q_L$ enters simultaneously in the quantities of Eqs.~(\ref{DefDeltat}) 
and (\ref{DefDeltab}). The values listed in Tab.~\ref{TabModels} hold simultaneously for the top and bottom sector, with an extra freedom of 
a global sign difference between the two. In what follows, the notation $\cM_i^+$ has been adopted for the same charge case, defined by 
$\Delta_\psi=\Delta_{\psi'}$, $\Delta_\chi=\Delta_{\chi'}$ and $\Delta_t=\Delta_b$, while $\cM_i^-$ for the opposite charge case, where 
$\Delta_\psi=-\Delta_{\psi'}$, $\Delta_\chi=-\Delta_{\chi'}$ and $\Delta_t=-\Delta_b$. The explicit charge assignment for each model can be read in Tab.~\ref{FullChargesAllModels}.

The scalar potential associated to all the models listed in Tab.~\ref{TabModels} has already been studied in Ref.~\cite{Merlo:2017sun}, 
together with the phenomenology associated to the exotic fermions and scalar fields. As a consequence, the next section will only focus 
on the ALP phenomenology.

%
%%%%%%%%%%%%%%%%%%%%%%%       III        %%%%%%%%%%%%%%%%%%%%%
%
\boldmath
\section{The ALP Phenomenology}
\unboldmath
\label{Sect:ALP}

Performing fermion field redefinitions, the Lagrangian in Eq.~(\ref{FermionLagrangian}) can be rewritten such that the axion or ALP has only derivative couplings with fermions. In particular, these models predict that the axion or ALP couples to both top and bottom quarks: these interactions can be written as
\be
\sL_a\supset -c_{a\psi\psi'}\dfrac{\derp_\mu a}{2f_a}\bar\psi\gamma^\mu\gamma_5\psi'\,,
\ee
where the couplings $c_{a\psi\psi'}$ depends on the specific model considered and can be read in Tab.~\ref{TableCoefficientTopBottom}.

\begin{table}[h!]
\hspace{-5mm}
\begin{tabular}{l|c|c|c|}
		  & $c_{att}$ & $c_{abb}$ \\
\hline
& & \\[-3mm]
$\cM_1^+$ & 
$2 n_s$ & 
$2 n_s$ 
\\
$\cM_1^-$ & 
$2 n_s$ & 
$-2 n_s$ 
\\[1mm]
\hline
& & \\[-3mm]
$\cM_2^+$ & 
$3 n_s$ & 
$3 n_s$ 
\\
$\cM_2^-$ & 
$3 n_s$ & 
$-3 n_s$ 
\\[1mm]
\hline
& & \\[-3mm]
$\cM_3^+$ & 
$4 n_s$ & 
$4 n_s$ 
\\
$\cM_3^-$ & 
$4 n_s$ & 
$-4 n_s$ 
\\[1mm]
\hline
& & \\[-3mm]
$\cM_4^+$ & 
$3 n_s$ & 
$3 n_s$ 
\\
$\cM_4^-$ & 
$3 n_s$ & 
$-3 n_s$ 
\\[1mm]
\hline
& & \\[-3mm]
$\cM_5^+$ & 
$2 n_s$ & 
$2 n_s$ 
\\
$\cM_5^-$ & 
$2 n_s$ & 
$-2 n_s$ 
\\[1mm]
\hline
& & \\[-3mm]
$\cM_6^+$ & 
$ n_s$ & 
$ n_s$ 
\\
$\cM_6^-$ & 
$ n_s$ & 
$- n_s$ 
\\[1mm]
\hline
& & \\[-3mm]
$\cM_7^+$ & 
$ n_s$ & 
$ n_s$ 
\\
$\cM_7^-$ & 
$ n_s$ & 
$- n_s$ 
\\
\hline
\end{tabular}
\caption{\em Values of the coefficients $c_{a\psi\psi'}$ in terms of the charge $n_s$ for the top and bottom quarks.}
\label{TableCoefficientTopBottom}
\end{table}

Moreover, at the quantum level, the derivative of the axial current is non-vanishing, giving rise to the following effective axion-gauge boson couplings: in the physical basis for the gauge bosons,
\be
\begin{aligned}
\delta\sL^\text{eff}_a\supset
&-\dfrac{\alpha_s}{8\pi}\,c_{agg}\,\dfrac{a}{f_a}\, G^a_{\mu\nu}\widetilde{G}^{a\mu\nu}
-\dfrac{\alpha_{em}}{8\pi}\,c_{a\gamma\gamma}\,\dfrac{a}{f_a}\,F_{\mu\nu}\widetilde{F}^{\mu\nu}+\\
&-\dfrac{\alpha_{em}}{8\pi}\,c_{aZZ}\,\dfrac{a}{f_a}\,Z_{\mu\nu}\widetilde{Z}^{\mu\nu}
-\dfrac{\alpha_{em}}{8\pi}\,c_{a\gamma Z}\,\dfrac{a}{f_a}\,F_{\mu\nu}\widetilde{Z}^{\mu\nu}+\\
&-\dfrac{\alpha_{em}}{8\pi}\,c_{aWW}\,\dfrac{a}{f_a}\,W^+_{\mu\nu}\widetilde{W}^{-\mu\nu}\,,
\end{aligned}
\ee
where $\widetilde{X}^{\mu\nu}\equiv \epsilon^{\mu\nu\rho\sigma}X_{\rho\sigma}/2$ and the convention $\epsilon_{1230}=+1$ is used. The mass independent anomaly contributions to the coefficients $c_{ai}$ are explicitly reported in App.~\ref{AxionLagrangian}, in terms of the PQ fermionic charges, while in Tab.~\ref{TableCoefficient} the anomalous coefficients for the seven models summarised in Tab.~\ref{TabModels} are listed.\footnote{Only one generation of SM fermions has been considered here, consistently with the formulation of the AML$\s$M presented in the previous section. Once extending this study to the realistic case of three generations, the values reported in Tab.~\ref{TableCoefficient} has to be modified: for example, assuming that the same charges will be adopted for all the fermion generations, the numerical values in the table would have to be multiplied by a factor 3.} These coefficients include the contributions of all the fermions that do couple with $a$.

\begin{table}[h!]
\hspace{-5mm}
\begin{tabular}{l|c|c|c|c|c|c|}
		  & $c_{agg}$ & $c_{a\gamma\gamma}$ & $c_{aZZ}$ & $c_{a\gamma Z}$ & $c_{aWW}$ \\
\hline
& & & & & \\[-2mm]
$\cM_1^+$ & 
$-2n_s$ & 
$-\frac{10}{3} n_s$ & 
$-\frac13n_s t^2_\theta-\frac{3n_s}{t^2_\theta}$ & 
$\frac23n_s t_\theta-\frac{6n_s}{t_\theta}$ & 
$-\frac{6 n_s}{s_\theta^2}$
\\
$\cM_1^-$ & 
$0$ & 
$-2n_s$ &
$-2n_s t^2_\theta$ & 
$4n_s t_\theta$ & 
$0$
\\[1mm]
\hline
& & & & & \\[-3mm]
$\cM_2^+$ & 
$-4n_s$ & 
$-\frac{20}{3} n_s$ & 
$-\frac{13}{6} n_s t^2_\theta-\frac{9n_s}{2t^2_\theta}$ & 
$\frac{13}{3} n_s t_\theta-\frac{9n_s}{t_\theta}$ & 
$-\frac{9 n_s}{s_\theta^2}$
\\
$\cM_2^-$ & 
$0$ & 
$-4n_s$ & 
$-4n_s t^2_\theta$ &  
$8n_s t_\theta$ & 
$0$
\\[1mm]
\hline
& & & & & \\[-3mm]
$\cM_3^+$ & 
$4n_s$ & 
$\frac{92}{3} n_s$ & 
$\frac{74}{3} n_s t^2_\theta+\frac{6n_s}{t_\theta^2}$ & 
$-\frac{148}{3} n_s t_\theta+\frac{12n_s}{t_\theta}$ & 
$\frac{12n_s}{s^2_\theta}$
\\
$\cM_3^-$ & 
$0$ & 
$4n_s$ & 
$4 n_s t^2_\theta$&  
$-8 n_s t_\theta$  & 
$0$
\\[1mm]
\hline
& & & & & \\[-3mm]
$\cM_4^+$ & 
$6n_s$ & 
$34 n_s$ & 
$\frac{53}{2} n_s t^2_\theta + \frac{15 n_s}{2t^2_\theta}$ & 
$-53n_s t_\theta+ \frac{15 n_s}{t_\theta}$ & 
$\frac{15 n_s}{s^2_\theta}$
\\
$\cM_4^-$ & 
$0$ & 
$6n_s$ &
$6 n_s t^2_\theta$ &  
$-12 n_s t_\theta$  & 
$0$
\\[1mm]
\hline
& & & & & \\[-3mm]
$\cM_5^+$ & 
$8n_s$ & 
$\frac{112}{3} n_s$ & 
$\frac{85}{3} n_s t^2_\theta+ \frac{9 n_s}{t^2_\theta}$ & 
$-\frac{170}{3} n_s t_\theta+ \frac{18 n_s}{t_\theta}$ & 
$\frac{18 n_s}{s^2_\theta}$
\\
$\cM_5^-$ & 
$0$ & 
$8n_s$ & 
$8n_s t^2_\theta$ &  
$-16n_s t_\theta$  & 
$0$
\\[1mm]
\hline
& & & & & \\[-3mm]
$\cM_6^+$ & 
$-10n_s$ & 
$-\frac{122}{3} n_s$ & 
$-\frac{163}{6} n_s t^2_\theta- \frac{27 n_s}{2t^2_\theta}$ & 
$\frac{163}{3} n_s t_\theta- \frac{27 n_s}{t_\theta}$ & 
$-\frac{27 n_s}{s^2_\theta}$
\\
$\cM_6^-$ & 
$0$ &  
$-10n_s$ & 
$-10 n_s t^2_\theta$ &  
$20 n_s t_\theta$  & 
$0$
\\[1mm]
\hline
& & & & & \\[-3mm]
$\cM_7^+$ & 
$0$ & 
$0$ & 
$\frac32 n_s t^2_\theta- \frac{3 n_s}{2t^2_\theta}$ & 
$-3 n_s t_\theta- \frac{3 n_s}{t_\theta}$  & 
$-\frac{3 n_s}{s^2_\theta}$
\\
$\cM_7^-$ & 
$0$ & 
$0$ & 
$0$ & 
$0$  & 
$0$
\\
\hline
\end{tabular}
\caption{\em Values of the coefficients $c_{ai}$ in terms of the charge $n_s$. $t_\theta$ and $s_\theta$ stand for the tangent and 
the sine of the Weinberg angle respectively.}
\label{TableCoefficient}
\end{table}

It is now possible to discuss the phenomenological features of the seven AML$\s$M constructions presented. Firstly, all models, but $\cM_7^\pm$, have a non-vanishing coupling between the ALP and two photons. As a consequence, the strong bound present on this coupling - reported in Eq.~(\ref{faBound}) - translates into a constraint on the scale $f_a$ that should be much larger than the EW scale, introducing a strong Hierarchy problem in the scalar potential (tree and loop level~\cite{deGouvea:2014xba}). In order to avoid this 
fine-tuning problem, $m_a\gtrsim0.1\GeV$ has to be considered for all the models $\cM_{1-6}^\pm$. As a drawback, none of these ALP models provide a solution to the Strong CP problem: such a large mass would correspond to an explicit breaking of the shift symmetry, 
perturbing the QCD potential and preventing the classical solution of the QCD axion models~\cite{Peccei:1977hh,Wilczek:1977pj,
Weinberg:1977ma,Shifman:1979if,Kim:1979if,Zhitnitsky:1980tq,Dine:1981rt}.

On the other hand, having $f_a$ in the TeV region opens the possibility of direct searches of ALP signatures at present and future experimental facilities~\cite{Merlo:2017sun,Bauer:2017ris,Arias-Aragon:2017eww}. An ALP with mass $m_a\sim1\GeV$ will be considered in the following as an illustration. 

For an ALP with a mass in the GeV region several constraints are present on its couplings to gauge bosons. Assuming that the ALP does not decay within the detector and therefore is treated as missing energy in data analyses, there are bounds from collider searches. In particular LEP data~\cite{Acciarri:1994gb,Anashkin:1999da} has been used to constrain ALP coupling to two photons~\cite{Mimasu:2014nea} once the axion is produced through a virtual photon: the corresponding bound on the scale $f_a$ reads
\be
\dfrac{f_a}{|c_{a\gamma\gamma}|}\gtrsim1\GeV\,.
\label{Cafotfot}
\ee
This bound may be improved by two order of magnitudes with dedicated analyses based on data from BaBar and from Belle-II~\cite{Mimasu:2014nea,Izaguirre:2016dfi,Dolan:2017osp}. Moreover, a similar sensitivity may be obtained considering the $\Upsilon(nS)\to\gamma+\text{inv.}$ decay~\cite{Dolan:2017osp,Merlo:2019anv}.

Studies on mono-$W$ and mono-$Z$ present LHC data~\cite{Brivio:2017ije} lead to 
\be
\dfrac{f_a}{|c_{aWW}|}\gtrsim0.7\GeV\,,\qquad
\dfrac{f_a}{|c_{aZZ}|}\gtrsim1.4\GeV\,, 
\label{boundaWW}
\ee
while LEP data~\cite{Acciarri:1994gb,Anashkin:1999da} on the radiative $Z$ decays has been used to infer a bound on  $a\gamma Z$ one~\cite{Dolan:2017osp}:
\be
\dfrac{f_a}{|c_{a\gamma Z}|}\gtrsim18\GeV\,.
\label{boundagammaZ}
\ee
Future LHC sensitivity prospects on mono-$W$ and mono-$Z$ considering an integrated luminosity of $3000\fb^{-1}$ improve the first two bounds of an order of magnitude~\cite{Brivio:2017ije}.

On the other side, rare meson decays provide strong constraints on the ALP coupling to two $W$'s. In the case of an invisible ALP, the most stringent bounds arise from Belle limits on $\mathcal{B}(B\to K\nu\bar{\nu})$~\cite{Grygier:2017tzo}. By assuming that only $c_{aWW}$ contributes, it leads to
\be
\dfrac{f_a}{|c_{aWW}|}\gtrsim10\GeV\,.
\label{CaWWstable}
\ee
Belle-II expected sensitivity improves this bound of approximately one order of magnitude~\cite{Izaguirre:2016dfi,Cunliffe:2017cox}.

Finally, considering the ALP coupling to top and bottom quarks, $B$ and $\Upsilon$ decays provide interesting bounds. Once considering that $B^+\to K^+a$ proceeds only via a loop diagram containing the $c_{att}$ coupling, the bound that can be extracted from Belle data~\cite{Grygier:2017tzo} for $m_a\approx 1\GeV$ reads~\cite{Gavela:2019wzg}
\be
\dfrac{f_a}{|c_{att}|}\gtrsim 200 \TeV\,,
\label{Cattstable}
\ee
while Belle-II may improve this bound of a factor of 5. In general, both $c_{aWW}$ and $c_{att}$ contribute to this decay and may exist part of the parameter space where a cancellation take place, relaxing the bounds in Eqs.~(\ref{CaWWstable}) and (\ref{Cattstable}). As discussed in Ref.~\cite{Gavela:2019wzg}, this cancellation is possible only if both $c_{aWW}$ and $c_{att}$ are loop-induced: this is not the case in the models discussed here, where $c_{att}$ is at tree-level.

Finally, data from BaBar~\cite{delAmoSanchez:2010ac,Aubert:2008as} and Belle~\cite{Seong:2018gut} on $\Upsilon(ns)\to\gamma+\text{inv.}$ put bounds on $f_a/|c_{abb}|$, but they are sub-dominant with respect to the previous bound from $B$ decays, reaching a sensitivity of a few TeV~\cite{Merlo:2019anv}: for $10\keV \lesssim m_a\lesssim5\GeV$,
\be
\dfrac{f_a}{|c_{abb}|}\gtrsim2.5\TeV\,,
\label{Cabbstable}
\ee
Also for $\Upsilon(ns)\to\gamma+\text{inv.}$, in general, the branching ratio would depend on both $c_{abb}$ and $c_{a\gamma\gamma}$: however, in the models considered here, $c_{a\gamma\gamma}$ is weighed by loop factors and then its contribution is negligible with respect to the one proportional to $c_{abb}$ (for the generic analysis see Ref.~\cite{Merlo:2019anv}). For heavier ALPs, there are only very weak bounds from colliders or B-factories, that would allow $f_a/|c_{abb}|$ to be in the TeV range. On the other side, for lighter masses, $m_a \lesssim 10\keV$, much stronger bounds from stellar cooling data~\cite{Feng:1997tn} are obtained:
\be
\begin{aligned}
\dfrac{f_a}{|c_{att}|}&\gtrsim1.2\times 10^6\TeV\qquad\text{for the top}\\
\dfrac{f_a}{|c_{abb}|}&\gtrsim6.1\times 10^2\TeV\qquad\text{for the bottom} \,.
\end{aligned}
\ee
These constraints have been derived translating the existing bounds on axion coupling to electrons into constraints on the axion emission occurring via a top or bottom loop.\\

When the ALP decays within the detector, other observables need to be considered. Focussing on the radiative ALP decay, LEP data~\cite{Acciarri:1994gb,Anashkin:1999da} on the radiative $Z$ decays can again be used to infer a bound on $a\gamma Z$ coupling~\cite{Dolan:2017osp}:
\be
\dfrac{f_a}{|c_{a\gamma Z}|}\gtrsim1.8\GeV\,,
\label{ALPDecaying1}
\ee
under the assumption that $\mathcal{B}(a\to\gamma\gamma)=1$.
These bounds may be improved by two order of magnitude with dedicated analyses both at B-factories and at LHC~\cite{Mimasu:2014nea,Izaguirre:2016dfi,Dolan:2017osp}. 

Although $a$ decays dominantly into photons in the models considered here ($c_{a\gamma\gamma}> c_{agg}$), the coupling with gluons can provide interesting phenomenology and can be bounded considering the BaBar results on the branching ratio of $\Upsilon(2s,3s)\to\gamma a(\to jj)$~\cite{Lees:2011wb}: for $m_a=1\GeV$,
\be
\dfrac{f_a}{|c_{agg}|}\gtrsim 80\GeV\,.
\label{ALPDecaying2}
\ee
This bound is expected to reach values of $0.2\TeV$ at Belle-II~\cite{CidVidal:2018blh}.

Finally, considering the ALP coupling with bottom quarks, data on $b\to sg$ or $b\to sq\ov{q}$ from CLEO collaboration~\cite{Coan:1997ye} allows to put bound on $f_a/|c_{abb}|$~\cite{Dolan:2014ska}: for $0.4\GeV\lesssim m_a\lesssim4.8\GeV$,
\be
\dfrac{f_a}{|c_{abb}|}\gtrsim2\TeV\,.
\label{ALPDecaying3}
\ee

On the other side, $B^\pm\to K^\pm\, a(\to2\gamma)$ decay, that could be studied at Belle-II, may be extremely useful to improve on these bound and will work as a test for the models presented here. Assuming that Belle-II reached a sensitivity of $10^{-6}$ on $\mathcal{B}(B\to K\gamma\gamma)$, strong bound can be inferred to $aWW$ and $att$ couplings: values of $f_a$ as large as $|c_{aWW}|\times60\GeV$ and $|c_{att}|\times 300\TeV$ could be probed.\\

To understand which bounds apply to the models listed above, the ALP decay length must be considered. It is  generically given by~\cite{Brivio:2017ije}
\be
d \sim \frac{10^{-2}}{c_{ai}^2} \left(\frac{\GeV}{m_a}\right)^4 \left(\frac{f_a}{\TeV}\right)^2 \left(\frac{|p_a|}{\GeV}\right)\,.
\ee
For an ALP of $m_a \sim 1\GeV$, $f_a\sim1\TeV$ and typical momentum $|p_a| \sim 100\GeV$, the traveling distance before decaying into two photons (the dominant channel as $|c_{a\gamma\gamma}|\gg |c_{agg}|$) is around $1/c_{a\gamma\gamma}^2\m$, that is in the interval $1\mm-0.1\m$ for $n_s=1$, depending on the specific value of $c_{a\gamma\gamma}$ reported in Tab.~\ref{TableCoefficient}. Therefore, all the ALPs described in the models $\cM_{1-6}^\pm$ decay within the detector and the bounds in Eqs.~(\ref{ALPDecaying1})--\eqref{ALPDecaying3} apply. The models $\cM_7^\pm$, instead, predict vanishing ALP couplings with both photons and gluons and therefore these ALPs are stable at tree level for masses up to $\sim 10\GeV$: for these models the bounds in Eqs.~(\ref{Cafotfot})-(\ref{Cabbstable}) apply.

Considering the explicit values of the axion coupling, taking $n_s=1$, the strongest bounds on $f_a$ for each model $\cM^\pm_{1-6}$ read as follow:
\be
\begin{aligned}
\cM_1^\pm&\longrightarrow f_a\gtrsim4\TeV\,,\qquad&
\cM_2^\pm&\longrightarrow f_a\gtrsim6\TeV\,,\\
\cM_3^\pm&\longrightarrow f_a\gtrsim8\TeV\,,\qquad&
\cM_4^\pm&\longrightarrow f_a\gtrsim6\TeV\,,\\
\cM_5^\pm&\longrightarrow f_a\gtrsim4\TeV\,,\qquad&
\cM_6^\pm&\longrightarrow f_a\gtrsim2\TeV\,.
\end{aligned}
\ee
For all these cases, $f_a$ can be in the TeV range, where the $SO(5)$ breaking mechanism is expected to occur.

\begin{figure}[h!]
\centering
\includegraphics[width=.8\linewidth]{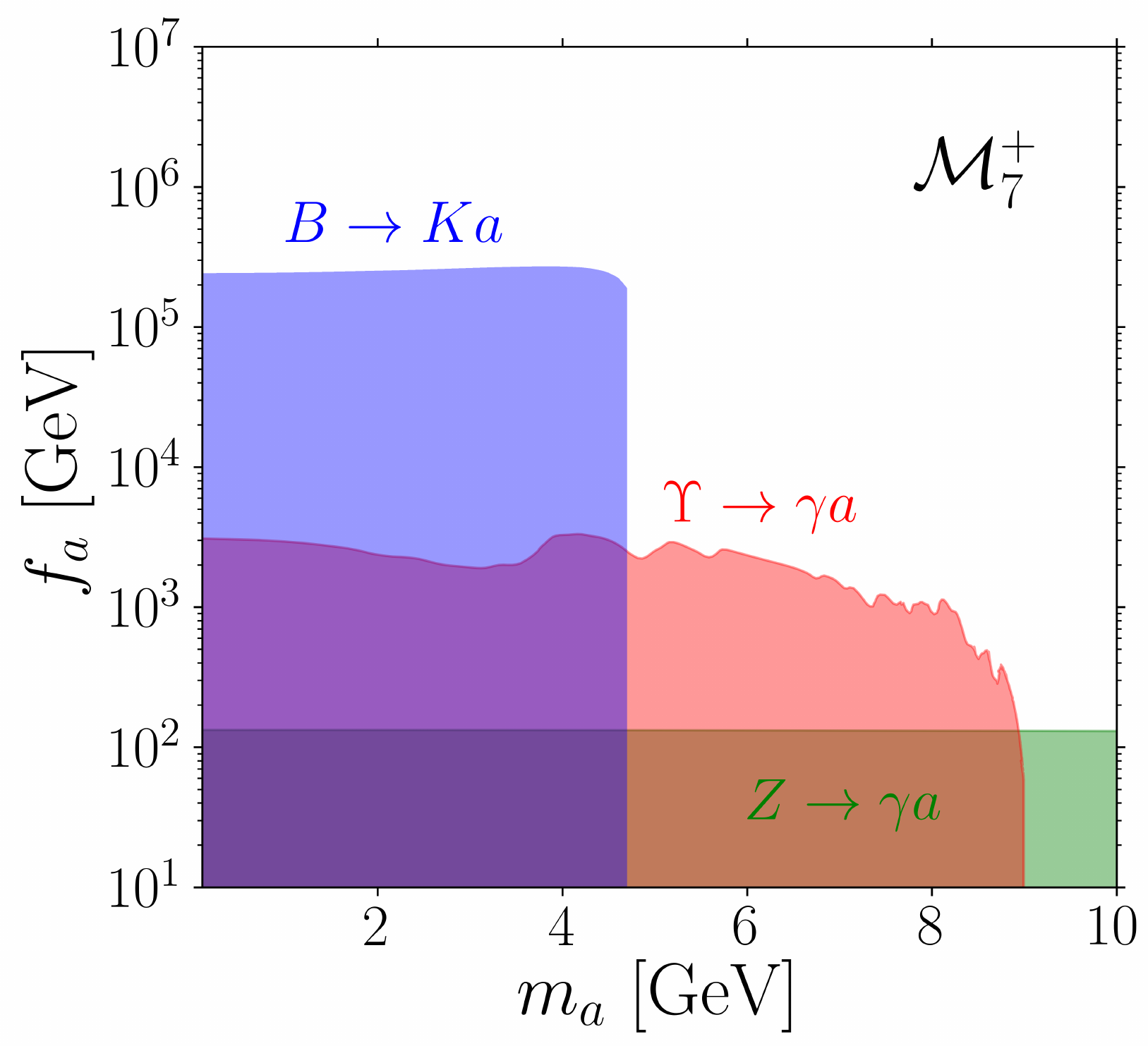}
\includegraphics[width=.8\linewidth]{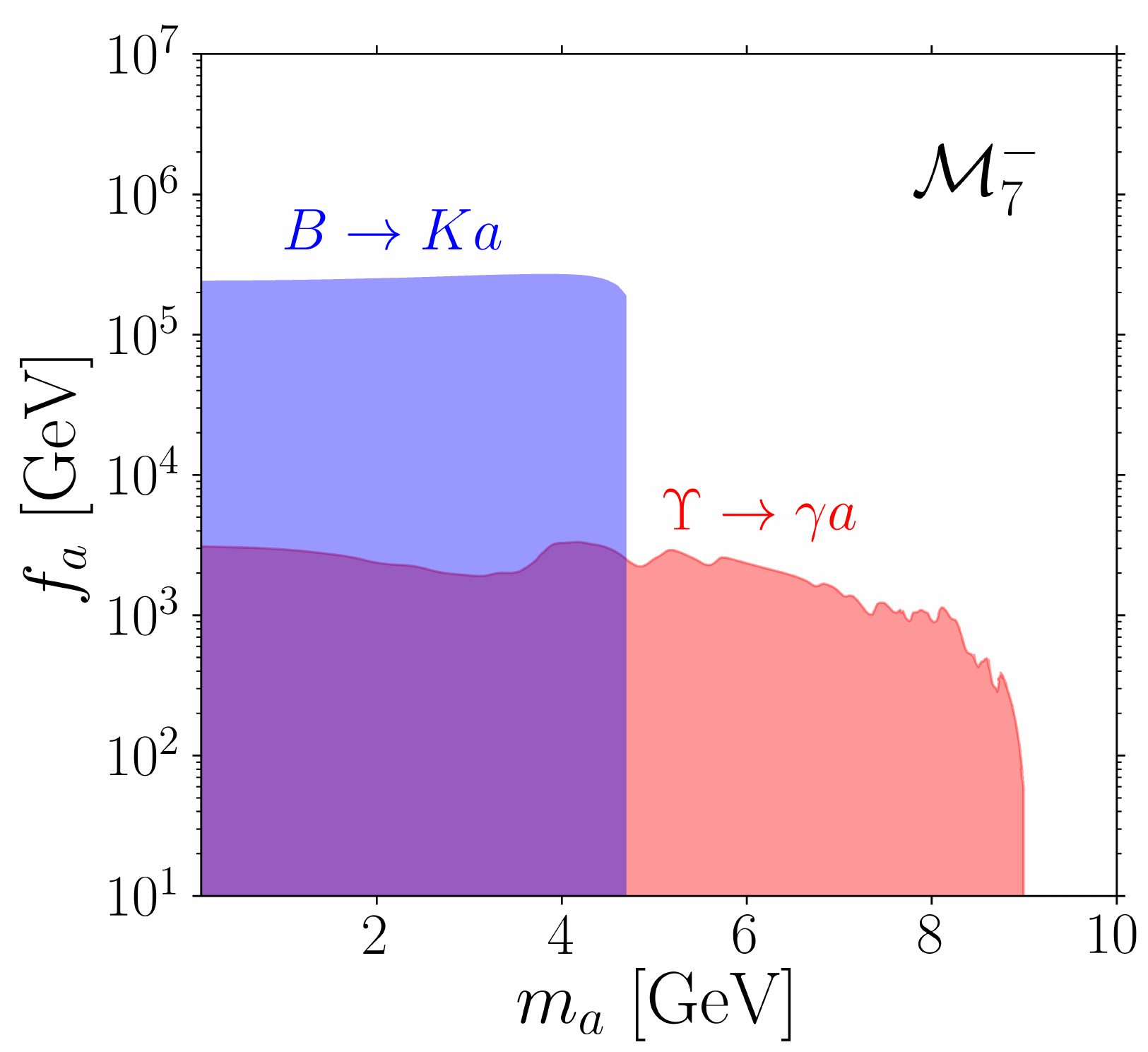}
\caption{\em Exclusions regions for the models $\cM_7^\pm$. The blue region corresponds to the bound from $B^+\to K^+a$ in Eq.~\eqref{Cattstable}. The red region to the bound from $\Upsilon(ns)\to\gamma+\text{inv.}$ in Eq.~\eqref{Cabbstable}. The green region to the collider bound from the radiative $Z$ decay in Eq.~\eqref{boundagammaZ}. The latter does not apply to the $\cM^-_7$ model because ALP only couples at tree level to bottom and top quarks.}
\label{Plots}
\end{figure}

On the other side, to summarise the most relevant constraints for $\cM_7^\pm$, the plots in Fig.~\ref{Plots} are shown. As can be seen, pretty strong constraints are present for $m_a\lesssim4.8\GeV$: in this case, $f_a\gtrsim200\TeV$ and therefore a mild tuning is present in the scalar potential of the models $\cM^\pm_7$. For masses larger than this value, but up to $~10\GeV$ the constraints are milder and $f_a\sim\mathcal{O}(1)\TeV$, avoiding any tuning in the scalar potential.\\

Finally, it is possible to comment {\it a posteriori} on the assumed value for $m_a$. In ALP models, corrections to the QCD potential are expected to be present such that the inverse proportionality between $m_a$ and $f_a$ gets broken: this is achieved with breaking terms at least of the order of $\Lambda_{QCD}\approx 100 \MeV$. As a consequence, the ALP mass is expected to be of the same order of these breaking terms and then larger than $\sim0.1\GeV$. On the order side, the ALP Lagrangian is written as an expansion in inverse powers of $f_a$, and for the EFT description to be meaningful, the ALP should be lighter than $f_a$. The benchmark scenario of $m_a=1\GeV$ enters in this range of values.

%
%%%%%%%%%%%%%%%%%%%%%%%       IV     %%%%%%%%%%%%%%
%
\section{Conclusions}
\label{Sect:Conc}

The strong bounds on axion/ALP couplings to photons and electrons imply very high values for the PQ breaking scale $f_a$. This represents the origin of a hierarchy problem present in most of the axion models: the quartic coupling in the potential between the Higgs field and the complex scalar field, associated to the axion, can be hardly ever prevented by symmetry arguments. As a consequence, avoiding fine-tuning among parameters, any other energy scale tends to be close to $f_a$. The AML$\s$M is a well defined and renormalisable framework where to address this problem. To delimitate the landscape of possible PQ charge assignments, three criteria have been imposed:  i) third generation SM fermion masses are generated at LO; ii) the $SO(5)$ and $\UPQ$ breaking scales coincide; iii) the model depends on the minimal possible number of parameters. Seven possible scenarios have been identified. 

For ALPs with masses larger than $10\MeV$, the aforementioned astrophysical bounds on couplings to photons and electrons are avoided. For concreteness, a mass of $1\GeV$ has been considered. This value is within the expected range of values for an ALP, that naturally has a mass larger than $\Lambda_{QCD}$ but smaller then $f_a$. In six of the models, the ALP decays within the detector and the most relevant bounds come from CLEO: they can be translated into constraints of the scale of the ALP, $f_a\gtrsim2\div8\TeV$. For the seventh model, the ALP does not decay and the strongest constraints arise from $B^+\to K^+a$ decay, the radiative $\Upsilon$ decay and collider analysis on radiative $Z$ decay: in this case, for ALP with masses up to $~4.5\GeV$, $f_a=f \sim 200\TeV$, while for larger masses up to $\sim10\GeV$ much weaker bounds are present and $f_a=f\sim\mathcal{O}(1)\TeV$.
Therefore, in all the realisations presented, the PQ and $SO(5)$ breaking scales can satisfy to $f_a=f \approx \mathcal{O}(1)\TeV$ and then these models have the possibilities to be tested soon both at colliders and at Belle-II (the only exception is $\cM^{\pm}_7$ and for ALP masses up to $\sim4.8\GeV$). In conclusion, these are testable and natural AML$\s$M, free from any fine-tuning in the scalar potential, and where the typical hierarchy problem that affects axion and ALP models is avoided.

%\newpage
%%%%%%%%%%%%%%%%%%%%%%%%%%%%%%%%%%%%%%%%%%%%%%%%%%%%%%%%%%%%%
\acknowledgments

The authors warmly thank I.~Brivio for comments and suggestions on the preliminary version of the paper, and F. Arias-Arag\'on, E. Fern\'andez-Mart\'inez, F. Mescia, S. Pokorski, and E. Stamou for useful discussions during the workout of the project.

The authors acknowledge partial financial support by the European Union's Horizon 2020 research and innovation programme under the Marie Sklodowska-Curie grant agreements No 690575 and No 674896. J.A.G and L.M. acknowledge partial financial support by the Spanish ``Agencia Estatal de Investigaci\'on'' (AEI) and the EU ``Fondo Europeo de Desarrollo Regional'' (FEDER) through the project FPA2016-78645-P, and through the Centro de excelencia Severo Ochoa Program under grant SEV-2016-0597. L.~M. acknowledges partial financial support by the Spanish MINECO through the ``Ram\'on y Cajal'' programme (RYC-2015-17173).

J.A.G. and L.M. thank the Physics and Astronomy department ``Galileo Galilei'' of the Padua University for hospitality during 
the development of this project. Furthermore, L.M. thanks the Kavli Institute for the Physics and Mathematics of the Universe for 
hospitality during the development of this project.

%%%%%%%%%%%%%%%%%%%%%%%%%%%%%%%%%%%%%%%%%%%%%%%%%%%%%%%%%%%%%
\appendix
%\newpage
\begin{widetext}

\section{Axion Lagrangian}
\label{APP:PQTrans}

The axion or ALP Lagrangian in the basis where axion-fermion couplings are only derivative is given by

\be
\sL_a =  -\frac{\partial_\mu a}{2f_a}\Big[
\Delta_{\psi} \bar{\psi}\gamma^\mu\psi+
\Delta_{\chi} \bar{\chi}\gamma^\mu\chi+
\Delta_{\psi'} \bar{\psi'}\gamma^\mu\psi'+
\Delta_{\chi'} \bar{\chi'}\gamma^\mu\chi'+
\Delta_{t} \bar{t}\gamma^\mu t+
\Delta_{b} \bar{b}\gamma^\mu b\Big]\,.
\ee

The axion or ALP couplings with gauge bosons arise due to the anomalous nature of the PQ symmetry. They can be read out in the following effective Lagrangian that encodes the traditional 1-loop contributions of all the fermions:
\be
\begin{aligned}
\delta\sL^\text{eff}_a\supset
&-\frac{\alpha_s}{8\pi}\Big[5\left(\Delta_\psi+\Delta_{\psi'}\right)+\Delta_\chi+\Delta_{\chi'}+\Delta_t+\Delta_b\Big]G_{\mu\nu}\widetilde{G}^{\mu\nu}+\\
&-\frac{\alpha_{em}}{8\pi}\Bigg[6\Delta_{\psi}\left(1+2\left(Y^2_{K}+Y^2_{Q}\right)+Y^2_{T_5}\right)+6\Delta_{\psi'}\left(1+2\left(Y^2_{K'}+Y^2_{Q'}\right)+Y^2_{B_5}\right)+\\
&\hspace{2cm}+6\left(\Delta_{\chi}Y^2_{T_1}+\Delta_{\chi'}Y^2_{B_1}\right)+6\left(\Delta_tY^2_{t_R}+\Delta_bY^2_{b_R}\right)\Bigg]F_{\mu\nu}\widetilde{F}^{\mu\nu}+\\
&-\frac{\alpha_{em}}{8\pi}\frac{6}{\sin^2\theta_W}\Big[2\left(\Delta_\psi+\Delta_{\psi'}\right) + \dfrac{\Delta_t+\Delta_b}{4} \Big]W^+_{\mu\nu}\widetilde{W}^{-\mu\nu}+\\
&-\frac{\alpha_{em}}{8\pi}\Bigg[6\Delta_{\psi}\left(\frac{1}{\tan^2\theta_W}+\tan^2\theta_W\left(2\left(Y^2_{K}+Y^2_{Q}\right)+Y^2_{T_5}\right)\right)+6\Delta_{\chi}\tan^2\theta_WY^2_{T_1}+\\
&\hspace{2cm}+6\Delta_{\psi'}\left(\frac{1}{\tan^2\theta_W}+\tan^2\theta_W\left(2\left(Y^2_{K'}+Y^2_{Q'}\right)+Y^2_{B_5}\right)\right)+6\Delta_{\chi'}\tan^2\theta_WY^2_{B_1}+\\
&\hspace{2cm}+3\dfrac{\Delta_t+\Delta_b}{4} \left(\frac{1}{\tan^2\theta_W} -\tan^2\theta_W\right)+6\tan^2\theta_W\left(\Delta_t Y_t^2+\Delta_b Y_b^2\right)\Bigg]Z_{\mu\nu}\widetilde{Z}^{\mu\nu}+\\
&-\frac{\alpha_{em}}{8\pi}\Bigg[12\Delta_{\psi}\left(\frac{1}{\tan\theta_W}-\tan\theta_W\left(2\left(Y^2_{K}+Y^2_{Q}\right)+Y^2_{T_5}\right)\right)-12\tan\theta_W\Delta_{\chi}Y^2_{T_1}+\\
&\hspace{2cm}+12\Delta_{\psi'}\left(\frac{1}{\tan\theta_W}-\tan\theta_W\left(2\left(Y^2_{K'}+Y^2_{Q'}\right)+Y^2_{B_5}\right)\right)-12\tan\theta_W\Delta_{\chi'}Y^2_{B_1}+\\
&\hspace{2cm}+3\dfrac{\Delta_t+\Delta_b}{2}\left(\frac{1}{\tan\theta_W}+\tan\theta_W\right)-12\tan\theta_W\left(\Delta_t Y_t^2+\Delta_b Y_b^2\right)\Bigg]F_{\mu\nu}\widetilde{Z}^{\mu\nu}\,.
\end{aligned}
\label{AxionLagrangian}
\ee
In the previous expression, $\Delta_\tf$ are defined in Eqs.~(\ref{DefDeltas}), (\ref{DefDeltat}) and (\ref{DefDeltab}), while $n_\tf$ is the PQ charge of the generic field $\tf$ that are reported for simplicity in Tab.~\ref{FullChargesAllModels}.\\

\begin{table}[h!]
{\hspace{-0.5cm}
\footnotesize
\begin{tabular}{l|c|}
\hline
& \\[-3mm]
$\cM_1^+$	& $n_{\psi_L}=n_{\chi_R}=n_{\psi_R}=n_{\chi_L}- n_s=n_{q_L}=n_{t_R}-2n_s$=$n_{\psi'_L}=n_{\chi'_R}=n_{\psi'_R}=n_{\chi'_L}- n_s=n_{b_R}-2n_s$ \\[2mm]
$\cM_1^-$		& $n_{\psi_L}=n_{\chi_R}=n_{\psi_R}=n_{\chi_L}- n_s=n_{q_L}=n_{t_R}-2n_s=n_{\psi'_L}=n_{\chi'_R}=n_{\psi'_R}=n_{\chi'_L}+ n_s=n_{b_R}+2n_s$\\[1mm]
\hline
& \\[-3mm]	
$\cM_2^+$	& $n_{\psi_L}=n_{\chi_R}=n_{\psi_R}=n_{\chi_L}- n_s=n_{q_L}+ n_s=n_{t_R}-2n_s=n_{\psi'_L}=n_{\chi'_R}=n_{\psi'_R}=n_{\chi'_L}- n_s=n_{q_L}+ n_s=n_{b_R}-2n_s$\\[2mm]	
$\cM_2^-$		& $n_{\psi_L}=n_{\chi_R}=n_{\psi_R}=n_{\chi_L}- n_s=n_{q_L}+ n_s=n_{t_R}-2n_s=n_{\psi'_L}+2n_s=n_{\chi'_R}+2n_s=n_{\psi'_R}+2n_s=n_{\chi'_L}+ 3n_s=n_{b_R}+4n_s$\\[1mm]
\hline
& \\[-3mm]
$\cM_3^+$	& $n_{\psi_L}=n_{\chi_R}=n_{\psi_R}+ n_s=n_{\chi_L}- n_s=n_{q_L}+2n_s=n_{t_R}-2n_s=n_{\psi'_L}=n_{\chi'_R}=n_{\psi'_R}+ n_s=n_{\chi'_L}- n_s=n_{b_R}-2n_s$\\[2mm]
$\cM_3^-$		& $n_{\psi_L}=n_{\chi_R}=n_{\psi_R}+ n_s=n_{\chi_L}- n_s=n_{q_L}+2n_s=n_{t_R}-2n_s=n_{\psi'_L}+4n_s=n_{\chi'_R}+4n_s=n_{\psi'_R}+ 3n_s=n_{\chi'_L}+ 5n_s=n_{b_R}+6n_s$\\[1mm]
\hline
& \\[-3mm]
$\cM_4^+$	& $n_{\psi_L}=n_{\chi_R}=n_{\psi_R}+ n_s=n_{\chi_L}- n_s=n_{q_L}+ n_s=n_{t_R}-2n_s=n_{\psi'_L}=n_{\chi'_R}=n_{\psi'_R}+ n_s=n_{\chi'_L}- n_s=n_{b_R}-2n_s$\\[2mm]
$\cM_4^-$		& $n_{\psi_L}=n_{\chi_R}=n_{\psi_R}+ n_s=n_{\chi_L}- n_s=n_{q_L}+ n_s=n_{t_R}-2n_s=n_{\psi'_L}+2n_s=n_{\chi_R}+2n_s=n_{\psi_R}+ n_s=n_{\chi_L}+3n_s=n_{b_R}+4n_s$\\[1mm]
\hline
& \\[-3mm]
$\cM_5^+$	& $n_{\psi_L}=n_{\chi_R}=n_{\psi_R}+ n_s=n_{\chi_L}- n_s=n_{q_L}=n_{t_R}-2n_s=n_{\psi'_L}=n_{\chi'_R}=n_{\psi'_R}+ n_s=n_{\chi'_L}- n_s=n_{b_R}-2n_s$\\[2mm]
$\cM_5^-$		& $n_{\psi_L}=n_{\chi_R}=n_{\psi_R}+ n_s=n_{\chi_L}- n_s=n_{q_L}=n_{t_R}-2n_s=n_{\psi'_L}=n_{\chi'_R}=n_{\psi'_R}- n_s=n_{\chi'_L}+n_s=n_{b_R}+2n_s$\\[1mm]
\hline
& \\[-3mm]
$\cM_6^+$	& $n_{\psi_L}=n_{\chi_R}=n_{\psi_R}- n_s=n_{\chi_L}- n_s=n_{q_L}- n_s=n_{t_R}-2n_s=n_{\psi'_L}=n_{\chi'_R}=n_{\psi'_R}- n_s=n_{\chi'_L}- n_s=n_{b_R}-2n_s$\\[2mm]
$\cM_6^-$		& $n_{\psi_L}=n_{\chi_R}=n_{\psi_R}- n_s=n_{\chi_L}- n_s=n_{q_L}- n_s=n_{t_R}-2n_s=n_{\psi'_L}-2n_s=n_{\chi'_R}-2n_s=n_{\psi'_R}- n_s=n_{\chi'_L}- n_s=n_{b_R}$\\[1mm]
\hline
& \\[-3mm]
$\cM_7^+$	& $n_{\psi_L}=n_{\chi_R}=n_{\psi_R}=n_{\chi_L}- n_s=n_{q_L}- n_s=n_{t_R}-2n_s=n_{\psi'_L}=n_{\chi'_R}=n_{\psi'_R}=n_{\chi'_L}- n_s=n_{b_R}-2n_s$\\[2mm]
$\cM_7^-$		& $n_{\psi_L}=n_{\chi_R}=n_{\psi_R}=n_{\chi_L}- n_s=n_{q_L}- n_s=n_{t_R}-2n_s=n_{\psi'_L}-2n_s=n_{\chi'_R}-2n_s=n_{\psi'_R}-2n_s=n_{\chi'_L}- n_s=n_{b_R}$\\[1mm]
\hline	
\end{tabular}
\caption{\em Definition of all the models in terms of the PQ charges of the fields as a function of $n_s$.
\label{FullChargesAllModels}}}
\end{table}
\end{widetext}

%%%%%%%%%%%%%%%%%%%%%%%%%%%%%%%%%%%%%%%%%%%%%%%%%%%%%%%%%%%%%
%\bibliography{biblio}{}
%\bibliographystyle{BiblioStyleLetter}

\providecommand{\href}[2]{#2}\begingroup\raggedright\endgroup

%%%%%%%%%%%%%%%%%%%%%%%%%%%%%%%%%%%%%%%%%%%%%%%%%%%%%%%%%%%%%
\end{document}